\documentclass[twocolumn,aps,showpacs,amsfonts,amsmath,amssymb,superscriptaddress,floatfix]
{revtex4}
\usepackage[dvips]{graphicx}

\newcommand{\be}{\begin{equation}}
\newcommand{\ee}{\end{equation}}
\newcommand{\affA}{%
     Communications Research Laboratory,
     Koganei, Tokyo 184-8795, Japan}
\newcommand{\affB}{%
     CREST, Japan Science and Technology Corporation}
\newcommand{\affC}{%
     Advanced Research Laboratory, Hitachi Ltd.,
     Hatoyama, Saitama 350-03, Japan}

\begin{document}
\title{Practical scheme for the optimal measuremnet 
in quantum interferometric devices}
\author{Masahiro Takeoka}
\affiliation{\affA}
\affiliation{\affB}
\author{Masashi Ban}
\affiliation{\affC}
\author{Masahide Sasaki}
\affiliation{\affA}
\affiliation{\affB}
%\affiliation{\affD}
%\email{e-mail:psasaki@crl.go.jp}
%
\begin{abstract}

We apply a Kennedy-type detection scheme, 
which was originally proposed for a binary communications system,
to interferometric sensing devices.
We show that the minimum detectable perturbation 
of the proposed system 
reaches the ultimate precision bound 
which is predicted by Neyman-Pearson hypothesis testing.
To provide concrete examples, we apply our interferometric scheme to 
phase shift detection by using coherent and 
squeezed probe fields.

\end{abstract}
\pacs{PACS numbers:03.65.Ta, 42.50.Dv, 07.60.Ly}
% 03.65.Ta Foundations of quantum mechanics; measurement theory
% 42.50.-p Quantum optics
% 07.60.Ly Interferometers  
\date{\today}
\maketitle

It is well known that the ultimate sensing precision of 
interferometric devices is limited by 
the quantum mechanical properties of the probing field
\cite{Lane93,Paris97,D'Ariano02}.
Precision limit analysis has conventionally been 
studied in the context of 
a quantum estimation problem \cite{Lane93}.
The problem was also recently treated as 
a binary decision problem 
based on the Neyman-Pearson criterion
\cite{Paris97,D'Ariano02}.
This criterion is often applied to the problem of detecting
small, low-rate perturbations,
such as gravitational waves.
The precision limit can be determined by 
the discrimination ability of 
the original probe state $\hat{\rho}_0$ and 
the perturbed probe state $\hat{\rho}_1$.

Neyman-Pearson hypothesis testing 
\cite{NeymanPearson33}
is a strategy to maximize the detection probability $P_{11}$
for fixed false-alarm probability $P_{01}$,
where 
$P_{11}$ is 
the probability that 
one will infer the state as $\hat{\rho}_1$ correctly, and 
$P_{01}$ is 
the probability that 
one will infer the state as $\hat{\rho}_1$ 
when $\hat{\rho}_0$ is true.
Here we consider 
a small perturbation modeled by a unitary operator 
$\hat{U}_p(g)$, 
and the pure probe state 
$\hat{\rho}_0=|\psi_0\rangle\langle\psi_0|$.
The small parameter shift to be detected is given by $g$.
In this restricted case, 
the maximum detection probability has been analytically derived 
as \cite{Helstrom_QDET}
%%%%%%%%%%%%%%%%%%%%
\begin{equation}
\label{eq:P_11}
P_{11} = \left\{ 
\begin{array}{cc}
\left[ \sqrt{P_{01}\kappa} + \sqrt{(1-P_{01})(1-\kappa)} \right]^2 &
0 \le P_{01} \le \kappa \\
1 & \kappa \le P_{01} \le 1
\end{array}\right. ,
\end{equation}
%%%%%%%%%%%%%%%%%%%%
where $\kappa=|\langle \psi_0 | \psi_1 \rangle|^2 
= |\langle\psi_0|\hat{U}_p(g)|\psi_0\rangle|^2$
with the perturbed state $|\psi_1\rangle=\hat{U}_p(g)|\psi_0\rangle$.
This general result has been applied 
to derive minimum detectable perturbation $g_M$ \cite{Paris97}.
Since the minimum threshold for $P_{11}$ to detect perturbation 
is given by 
%%%%%%%%%%%%%%%%%%%%
\begin{equation}
\label{eq:threshold}
P_{11}(g_{\rm M};P_{01})=\frac{1}{2},
\end{equation}
%%%%%%%%%%%%%%%%%%%%
one can figure out the value of $g_{\rm M}$ 
for given probe states
from Eqs.~(\ref{eq:P_11}) and (\ref{eq:threshold}).

Although these analyses can be used to 
predict the bounds of 
ultimate precision limits 
for given probe states,
they tell us nothing about 
how to design optimal measurement devices 
in practice.
A practical measurement scheme has only been reported for 
a certain entangled probe field \cite{D'Ariano02}.
In this paper, 
we discuss the practical implementation of 
optimal measurement 
based on the so-called Kennedy scheme \cite{Kennedy73} 
which was originally proposed for semi-optimal detection strategy 
in terms of the average error probability 
for the binary phase-shift keyed coherent states 
$\{ |\alpha\rangle,\,|-\alpha\rangle \}$.
The scheme consists of the displacement operation 
$\hat{D}(\alpha)=\exp[\alpha\hat{a}^{\dagger}-\alpha^*\hat{a}]$
and the photodetection operation
$\{ \hat{\bf I}-|0\rangle\langle0|, |0\rangle\langle0| \}$
discriminating $|2\alpha\rangle$ and $|0\rangle$.
Since the signal $|0\rangle$ is perfectly projected into 
the second measurement operator, in principle, 
the total bit error rate performance is greatly enhanced. 
It is indeed semi-optimal to the criterion 
developed by Helstrom \cite{Helstrom_QDET}.

\begin{figure}
\begin{center}
\includegraphics[width=0.45\textwidth]{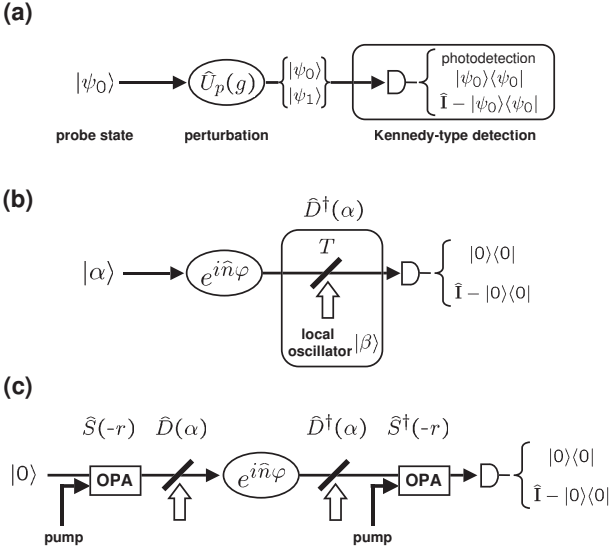}
\end{center}
\caption{\label{fig:Kennedy-scheme}
(a) Schematic illustration of Kennedy-type interferometer.
Phase shift detection by the Knnedy-type detector with 
(b) coherent probe field, and 
(c) squeezed probe field are also illustrated.
}
\end{figure}

We will now apply this concept 
to interferometric sensing devices. 
The outline of our scheme is 
in Fig.~\ref{fig:Kennedy-scheme}(a).
The set of measurement operators is given by the POVM 
%%%%%%%%%%%%%%%%%%%%
\begin{eqnarray}
\label{eq:KennedyPOVM}
\left\{ 
\begin{array}{l}
\hat{\Pi}_0 = |\psi_0\rangle \langle\psi_0| \\
\hat{\Pi}_1 = \hat{\bf I} - \hat{\Pi}_0
\end{array}
\right. .
\end{eqnarray}
%%%%%%%%%%%%%%%%%%%%
Since $\langle \psi_0 | \hat{\Pi}_0 | \psi_0 \rangle = 1$, 
the false-alarm probability $P_{01}$ is always zero.
The detection probability $P_{11}$ is given by 
%%%%%%%%%%%%%%%%%%%%
\begin{equation}
\label{eq:KennedyP_11}
P_{11} = 
\langle \psi_1 | (\hat{\bf I} - \hat{\Pi}_0) | \psi_1 \rangle 
%{\rm \,Tr}\, \left[ \rho_1 (\hat{\bf I} - \hat{\Pi}_0) \right]
= 1-\kappa.
\end{equation}
%%%%%%%%%%%%%%%%%%%%
Comparing these probabilities 
to the predicted bound in Eq.~(\ref{eq:P_11}), 
we can see that our scheme achieves
the optimal POVM for Neyman-Pearson hypothesis testing
where the false-alarm probability is zero, 
i.e. Eq.~(\ref{eq:KennedyP_11}) 
equals the $P_{11}(P_{01}=0)$ of Eq.~(\ref{eq:P_11}).
Obviously, the minimum detectable perturbation $g_{\rm M}$ 
derived from Eq.~(\ref{eq:KennedyP_11}) achieves 
the ultimate limit predicted by the Neyman-Pearson approach.

As a concrete example, let us now discuss an ordinary interferometer 
which detects small phase shifts given by 
the operator $\hat{U}_p=\exp(i\hat{n}\varphi)$.
Here, $\hat{n}$ is 
the photon number operator and $\varphi$ is the parameter for 
small phase shift.
Needless to say, this is the most conventional interferometric device 
commonly used 
in various sensing applications.

First, let us consider 
a coherent state $|\psi_0\rangle = |\alpha\rangle$
as a probe field quantum state.
Without loss of generality, we can assume that $\alpha$ is real.
Figure~\ref{fig:Kennedy-scheme}(b) has a possible setup.
The coherent probe beam is incident on the blackbox in which 
phase shifting occurs with very small probability.
At the receiving side, the probe field goes through 
a beamsplitter with the power transmission $T$, and 
it is detected by a photodetector 
to discriminate whether the field 
includes zero or non-zero photons.
A strong local oscillator $|\beta\rangle$ interferes
with the probe field 
from the other port of the beamsplitter.
As is well known \cite{Banaszek99},
in the limits $T \to 1$ and $\beta \to \infty$, 
the beamsplitter works as a displacement operator 
$\hat{D}(\sqrt{1-T}\beta)$.
For our purposes, the displacement has been carefully tuned to 
$\sqrt{1-T}\beta=-\alpha$.
The inference probabilities $P_{11}$ and $P_{01}$ can 
then be calculated as
%%%%%%%%%%%%%%%%%%%%
\begin{eqnarray}
\label{eq:P_11_coherent}
P_{11} & = & 1 - |\langle\alpha| e^{i\hat{n}\varphi} 
|\alpha\rangle|^2 \nonumber\\
& = & 1 - \exp [ - 2 \alpha^2 (1 - \cos\varphi) ],
\end{eqnarray}
%%%%%%%%%%%%%%%%%%%%
and $P_{01}=0$, respectively.
As discussed previously, this achieves the ultimate bound 
predicted by the Neyman-Pearson optimization procedure, 
which means that this simple setup provides the best measurement strategy 
for a coherent probe field. 
Expanding Eq.~(\ref{eq:P_11_coherent}) into the second order of $\varphi$ and 
adopting the minimum threshold condition Eq.~(\ref{eq:threshold}), 
we can find the minimum detectable phase shift $\varphi_{\rm M}^{coh}$ as 
%%%%%%%%%%%%%%%%%%%%
\begin{equation}
\label{eq:phi_coherent}
\varphi_{\rm M}^{coh} \approx \sqrt{\frac{\ln 2}{\langle n \rangle}}, 
\end{equation}
%%%%%%%%%%%%%%%%%%%%
where $\langle n \rangle =|\alpha|^2$ is the mean photon number of 
the probe field.

%%%%%%%%%%%%%%%%%%%%
%%%%%%%%%%%%%%%%%%%%

As a second example, let us consider 
a Kennedy scheme for a squeezed probe field, 
as illustrated in Fig.~\ref{fig:Kennedy-scheme}(c).
The probe field is
the so-called ``ideal squeezed state \cite{Mandel_OCandQO}'' 
, defined by 
$|\psi_0\rangle=\hat{D}(\alpha)\hat{S}(-r)|0\rangle$.
Here 
$\hat{S}(\zeta)=\exp[-\frac{1}{2}\zeta(\hat{a}^{\dagger \,2}-\hat{a}^2)]$ 
and 
$r$ is real and positive.
The detection system consists of a displacement beamsplitter 
$\hat{D}^{\dagger}(\alpha)$, 
a squeezer $\hat{S}^{\dagger}(-r)$, 
and a photodetector.
Two squeezers may be suitably realized by 
traveling wave optical parametric amplification
(OPA) \cite{Kim94}.
The overlap $\kappa$
is now given by
%%%%%%%%%%%%%%%%%%%%
\begin{eqnarray}
\label{eq:kappa_squeezed}
\kappa & = & 
|\langle0| \hat{S}^{\dagger}(-r) \hat{D}^{\dagger}(\alpha) 
e^{i\hat{n}\varphi} \hat{D}(\alpha) \hat{S}(-r) |0\rangle|^2 \nonumber\\
& = &
\frac{1}{\sqrt{\sigma_1 \sigma_2}} \exp \left[
- 2 e^{-2r} \alpha^2 \left( 1
- \frac{\cos\varphi}{\sigma_1 \sigma_2} \right. \right. \nonumber\\
& & \left. \left.
+ \frac{(e^{4r}-e^{-4r}) \sin^2 \varphi}{4\sigma_1 \sigma_2}
\right) \right],
\end{eqnarray}
%%%%%%%%%%%%%%%%%%%%
where
%%%%%%%%%%%%%%%%%%%%
\begin{equation}
\label{eq:sigma_12}
\sigma_{1,2} =  
\frac{1}{2} \left( e^{2r}(1 \mp \cos\varphi) 
+ e^{-2r}(1 \pm \cos\varphi) \right).
\end{equation}
%%%%%%%%%%%%%%%%%%%%
For small $\varphi$, $\kappa$ can be approximated to 
%%%%%%%%%%%%%%%%%%%%
\begin{equation}
\label{eq:kappa_squeezed_approx}
\kappa \approx \frac{1}{\sqrt{1+\sinh^2 2r \, \varphi^2}}
\exp \left[ - \frac{e^{2r}\alpha^2\varphi^2}{
1+\sinh^2 2r \, \varphi^2} \right]. 
\end{equation}
%%%%%%%%%%%%%%%%%%%%
Then, from Eqs.~(\ref{eq:threshold}), (\ref{eq:KennedyP_11}) 
and (\ref{eq:kappa_squeezed_approx}),
the minimum detectable phase shift $\varphi_{\rm M}^{sq}$ is given by 
%%%%%%%%%%%%%%%%%%%%
\begin{equation}
\label{eq:phi_squeezed}
\varphi_{\rm M}^{sq} \approx
\sqrt{\frac{1}{\sinh^2 2r} \left(
\frac{2e^{2r}\alpha^2}{\sinh^2 2r \, W(z)}-1 \right)},
\end{equation}
%%%%%%%%%%%%%%%%%%%%
where $W(x)$ is the product log function which is defined by 
the principal solution for $w$ in $x=w e^w$ 
and $z$ is 
%%%%%%%%%%%%%%%%%%%%
\begin{equation}
\label{eq:z}
z = \frac{e^{2r} \alpha^2}{2\sinh^2 2r} 
\exp \left[ \frac{2e^{2r}\alpha^2}{\sinh^2 2r} \right].
\end{equation}
%%%%%%%%%%%%%%%%%%%%

\begin{figure}
\begin{center}
\includegraphics[width=0.35\textwidth]{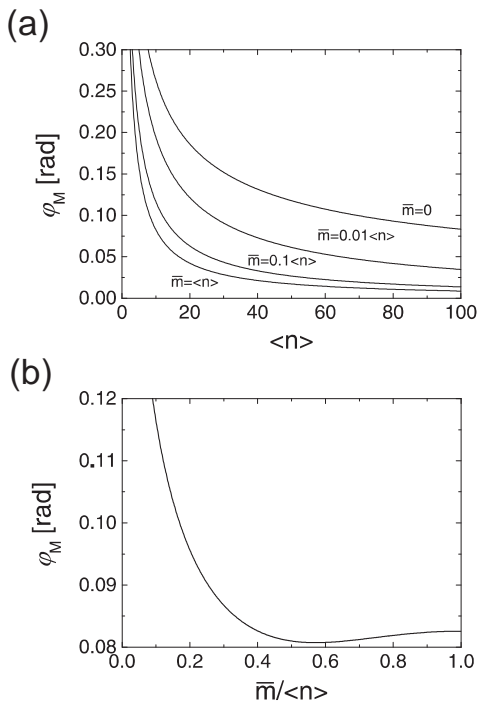}
\end{center}
\caption{\label{fig:Phi_Squeezed}
Minimum detectable phase shift $\varphi_{\rm M}$ as functions of 
(a) total photon number $\langle n \rangle$, and 
(b) ratio of squeezing power $\bar{m}/\langle n \rangle$ 
with $\langle n \rangle =10$.
}
\end{figure}

\begin{figure}
\begin{center}
\includegraphics[width=0.40\textwidth]{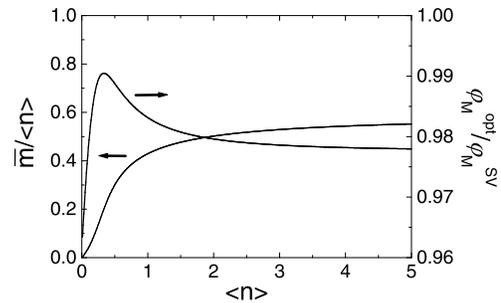}
\end{center}
\caption{\label{fig:Phi_Sq_Opt}
Optimal power distribution for coherent amplitude and squeezing 
in a probe field.
Minimum detectable phase shift $\varphi_{\rm M}^{\rm opt}$ for 
this optimized probe filed is also plotted.
}
\end{figure}

It is worth comparing performances by coherent 
and squeezed probe fields, under 
the power constraint condition 
$\langle n \rangle = \bar{n} + \bar{m} = |\alpha|^2 + \sinh^2 r$,
where $\langle n \rangle $, $\bar{n}$ and $\bar{m}$ are 
the total photon number, $\bar{n}=|\alpha|^2$ and $\bar{m}=\sinh^2 r$, 
respectively.
The minimum detectable phase shift $\varphi_{\rm M}$ for a given 
$\langle n \rangle$ is plotted in Fig.~\ref{fig:Phi_Squeezed}(a) with 
$\bar{m}=0$(coherent state), 
$0.01\langle n \rangle$, $0.1\langle n \rangle$ 
and $\langle n \rangle$(squeezed vacuum), 
while Fig.~\ref{fig:Phi_Squeezed}(b) shows 
the dependence of $\varphi_{\rm M}$
on the power distribution ratio $\bar{m}/\langle n \rangle$ 
for $\langle n \rangle=10$.
The latter clearly indicates that 
there is an optimal power distribution between 
$0<\bar{m}<\langle n \rangle$ to obatain 
the lowest precision limit $\varphi^{opt}_{\rm M}$.
Such $\varphi^{opt}_{\rm M}$ is plotted 
in Fig.~\ref{fig:Phi_Sq_Opt} 
with the corresponding power ratio $\bar{m}/\langle n \rangle$.
As a reference, $\varphi^{opt}_{\rm M}$ is normalized by 
the $\varphi_{\rm M}$ for the squeezed vacuum, 
which is denoted by $\varphi^{sv}_{\rm M}$.
The figure indicates that, in the limit of large $\langle n \rangle$, 
$\varphi^{opt}_{\rm M}$ asymptotically reaches about 98\% of 
$\varphi^{sv}_{\rm M}$ where the power distribution reaches about 
$\bar{m}/\langle n \rangle \approx 0.55$.

When the power for the probe field is fully used for squeezing, 
i.e., $\bar{m}=\langle n \rangle$ and $\bar{n}=0$, 
the $\varphi^{sq}_{\rm M}$ of Eq.~(\ref{eq:phi_squeezed}) 
can be simplified to 
%%%%%%%%%%%%%%%%%%%%
\begin{equation}
\label{eq:phi_m=<n>}
\varphi^{sv}_{\rm M} =
\frac{\sqrt{3}}{\sinh 2r} =
\sqrt{\frac{3}{4\langle n \rangle(\langle n \rangle+1)}}.
\end{equation}
%%%%%%%%%%%%%%%%%%%%
In the limit of large $\langle n \rangle$,
$\varphi_{\rm M}$ is proportional to $1/\langle n \rangle$.
This is similar to that of previous predictions 
\cite{Lane93,D'Ariano02}.
From Fig.~\ref{fig:Phi_Sq_Opt} and Eq.~(\ref{eq:phi_m=<n>}),
we can conclude that when $\langle n \rangle$ is enough large, 
$\varphi^{opt}_{\rm M}$ is also proportional to $1/\langle n \rangle$.
Nevertheless, our results clearly show that 
when we are only allowed to use 
extremely weak probe field, 
an optimization of the power distribution certainly improves
the precision limit compared to not only coherent state, 
but also the squeezed vacuum state.

On the other hand, from a practical point of view,
it is very difficult to prepare a squeezed state with large $\bar{m}$
while large coherent amplitude can easily be generated.
If we assume that $\bar{n} \gg \bar{m}$, 
i.e., $\bar{m}$ is significantly smaller than 
$1/\varphi_{\rm M}$, 
$\varphi^{sq}_{\rm M}$ can simply be calculated by
%%%%%%%%%%%%%%%%%%%%
\begin{equation}
\label{eq:phi_small_m}
\varphi^{sq}_{\rm M} = \frac{1}{e^r}
\sqrt{\frac{\ln 2}{\bar{n}}},
\end{equation}
%%%%%%%%%%%%%%%%%%%%
which indicates how squeezing enhances 
the precision limit of small phase shift detection 
in a bright squeezed probe field.

%%%%%%%%%%%%%%%%%%%%
%%%%%%%%%%%%%%%%%%%%

%%%%%%%%%%%%%%%%%%%%
%%%%%%%%%%%%%%%%%%%%

We also need to note that even though
our scheme can directly be applied 
to the two-mode squeezed state probing, 
it has no advantages compared to 
the single mode squeezed state.
This is because the advantage of using 
the two-mode squeezed state, i.e. entanglement, 
instead of the single-mode squeezed state 
for interferometric devices 
is understood as 
its stability against the technical phase fluctuations \cite{D'Ariano02}, 
while our scheme requires the use of local oscillators.
However, our scheme only requires devices 
that can presently be obtained and 
the photodetector restrictions are less severe 
than with the scheme utilizing entanglement, 
especially in the extremely weak probe field region,
which has the squeezing power that is currently available.
While the entanglement scheme assumes 
to detect the difference of the photon number between two modes 
 \cite{D'Ariano02}
which means detectors need to resolve the number of photons,
our scheme only requires a detector that can discriminate 
between zero or non-zero photons.
This kind of photodetection is possible by 
extending the current technology 
e.g., by using an avalanche photodiode (APD)
operating in the Geiger mode.
In practice, APDs are parametrized by quantum efficiency $\eta$
and dark current $I_d$ and the latter 
causes serious false-alarm probability and also 
decreases the detection probability.
Typical quantities for the best devices that are 
commericially available at present 
are $\eta \sim 80\%$ and $I_d \sim 50$ 
counts per second, for example.
We therefore need to pursue further quantitative improvements 
in detecting devices. 
Nevertheless, we believe that our scheme still represents 
a straightforward extension of current photodetection technology.

To summarize, we applied the concept of a Kennedy detection scheme, 
which has been studied in the field of communications theory, 
to interferometric sensing devices.
We showed that the ultimate precision of our physically realizable 
scheme reaches 
the ultimate precision bound predicted by 
Neyman-Pearson hypothesis testing.
It allows us to design concrete 
optimal detection apparatuses for various given probe sources,
e.g., coherent or squeezed states.
These are useful in various applications where 
very small signals must reliably be detected, especially 
in regions where only weak probe power is available.

\acknowledgements
The authors acknowledge Dr.~M.~Fujiwara and Dr.~J.~Mizuno for 
valuable comments and discussions.


\begin{references}

\bibitem{Lane93}
   A.~S.~Lane, S.~L.~Braunstein, and C.~M.~Caves, 
   Phys.\ Rev.\ A\,\textbf{47}, 1667 (1993). 

\bibitem{Paris97}	
   M.~G.~Paris,
   Phys.\ Lett.\ A\,\textbf{225}, 23 (1997). 

\bibitem{D'Ariano02}
   G.~M.~D'Ariano, M.~G.~A.~Paris, and P.~Perinotti,
   Phys.\ Rev.\ A\,\textbf{65}, 062106 (2002). 

\bibitem{NeymanPearson33}
   J.~Neyman and E.~Pearson, 
   Philos.\ Trans.\ R.\ Soc.\ London,\ Ser.\ A\,\textbf{231}, 289 (1933).

\bibitem{Kennedy73}
   R.~S.~Kennedy, 
   Research Laboratory of Electronics, MIT, 
   Quarterly Progress Report No.~108, 219 (1973).

\bibitem{Helstrom_QDET}
   C.~W.~Helstrom, {\it Quantum Detection and Estimation Theory}
   (Academic Press, New York, 1976).

\bibitem{Banaszek99}
   K. Banaszek and K. Wodkiewicz, 
   Phys. Rev. Lett.\,\textbf{82}, 2009 (1999).

\bibitem{Mandel_OCandQO}
    L.~Mandel and E.~Wolf, {\it Optical Coherence and Qunatum Optics}
    (Cambridge University Press, Cambridge, 1995).

\bibitem{Kim94}
   C.~Kim and P.~Kumar,
   Phys.\ Rev.\ Lett.\,\textbf{73}, 1605 (1994). 

\end{references}
\end{document}